\documentclass[aps,prx,twocolumn,floatfix,superscriptaddress,longbibliography,notitlepage]{revtex4-2}
\usepackage{braket}
\usepackage{amsmath}
\usepackage{amssymb}
\usepackage[utf8]{inputenc}
\usepackage[T1]{fontenc}
\usepackage{tikz}
\usetikzlibrary{arrows,shapes,calc,matrix}
\usepackage{upgreek} 

\usepackage{changes}
\setdeletedmarkup{\color{blue}\sout{#1}}
\usepackage{cancel}

\usepackage{mathrsfs}
\usepackage[normalem]{ulem}
\usepackage{graphicx, color, graphpap}
\usepackage{enumitem}
\usepackage{amsthm}
\usepackage{multirow}
\usepackage[colorlinks=true,citecolor=blue,linkcolor=magenta]{hyperref}
\usepackage[T1]{fontenc}
\usepackage{verbatim}
\usepackage{mathtools}
\usepackage{titlesec}

\long\def\ca#1\cb{} 

\newcommand{\abs}[2][]{#1| #2 #1|}

\newcommand{\ketbra}[2]{| \hspace{1pt} #1 \rangle \langle #2 \hspace{1pt} |}

\newcommand{\bramatket}[3]{\langle #1 \hspace{1pt} | #2 | \hspace{1pt} #3 \rangle}

\newcommand{\dya}[1]{\ket{#1}\!\bra{#1}}

\newcommand{\rank}{\text{rank}}

\newcommand{\Tr}{{\rm Tr}}

\renewcommand{\geq}{\geqslant}
\renewcommand{\leq}{\leqslant}

\newcommand*{\id}{\openone}

\renewcommand{\vec}[1]{\boldsymbol{#1}}  

\newcommand{\ad}{^\dagger}

\newtheorem{theorem}{Theorem}
\newtheorem{lemma}{Lemma}
\newtheorem{corollary}{Corollary}

\newtheorem*{theorem*}{Theorem}

\begin{document}
\title{No-go theorem for environment-assisted invariance in non-unitary dynamics}

\author{Akira Sone}
\email{akira.sone@umb.edu}
\affiliation{Department of Physics, University of Massachusetts, Boston, Massachusetts 02125, USA}
\affiliation{The NSF AI Institute for Artificial Intelligence and Fundamental Interactions}

\author{Akram Touil}
\affiliation{Theoretical Division, Los Alamos National Laboratory, Los Alamos, New Mexico 87545, USA}

\author{Kenji Maeda}
\affiliation{Department of Physics, University of Massachusetts, Boston, Massachusetts 02125, USA}

\author{Paola Cappellaro}
\affiliation{Department of Nuclear Science and Engineering, Massachusetts Institute of Technology, Cambridge, Massachusetts 02139, USA}
\affiliation{Research Laboratory of Electronics, Massachusetts Institute of Technology, Cambridge, Massachusetts 02139, USA}
\affiliation{Department of Physics, Massachusetts Institute of Technology, Cambridge, Massachusetts 02139, USA}

\author{Sebastian Deffner}
\affiliation{Department of Physics, University of Maryland, Baltimore County, Baltimore, MD 21250, USA}
\affiliation{Quantum Science Institute, University of Maryland, Baltimore County, Baltimore, MD 21250, USA}
\affiliation{National Quantum Laboratory, College Park, MD 20740, USA}

\begin{abstract}
We elucidate the requirements for quantum operations that achieve environment-assisted invariance (envariance), a symmetry of entanglement. While envariance has traditionally been studied within the framework of local unitary operations, we extend the analysis to consider non-unitary local operations. First, we investigate the conditions imposed on operators acting on pure bipartite entanglement to attain envariance. We show that the local operations must take a direct-sum form in their Kraus operator representations, establishing decoherence-free subspaces. Furthermore, we prove that this also holds for the multipartite scenario. As an immediate consequence, we demonstrate that environment-assisted shortcuts to adiabaticity cannot be achieved through non-unitary operations. In addition, we show that the static condition of the eternal black hole in AdS/CFT is violated when the CFTs are coupled to the external baths.
\end{abstract}

\maketitle

\section{Introduction}
\label{sec:intro}
Envariance, or environment-assisted invariance, is a fundamental concept in quantum mechanics, and particularly suited to the study of open quantum systems.  Originally introduced by Zurek to derive Born's rule~\cite{Zurek2003PRL,Zurek2003RMP,Zurek2005PRA,Zurek2009NatPhys,Zurek2011PRL,zurek2022quantum,Zurek_2025}, envariance characterizes a symmetry of quantum states with respect to unitary quantum maps. This is best illustrated by a simple example,
\begin{equation}
    \begin{array}{c}
         \begin{tikzpicture}[>=stealth,baseline,anchor=base,inner sep=0pt]
            \matrix (foil1) [matrix of math nodes,nodes={minimum height=0.1em}] {
                & {\color{blue}\ket{\uparrow}_{\mathcal{S}}} & \otimes & \ket{\uparrow}_{\mathcal{E}} &  & + &  & {\color{blue}\ket{\downarrow}_{\mathcal{S}}} & \otimes & \ket{\downarrow}_{\mathcal{E}} &  \xrightarrow{~ {\color{blue}U_{\mathcal{S}}}~} 
                {\color{blue}\ket{\downarrow}_{\mathcal{S}}}\otimes\ket{\uparrow}_{\mathcal{E}}+{\color{blue}\ket{\uparrow}_{\mathcal{S}}}\otimes\ket{\downarrow}_{\mathcal{E}}\\
            };
            \path[->] ($(foil1-1-2.north)+(0,1ex)$) edge[blue,bend left=45] ($(foil1-1-8.north)+(0,1ex)$);
            \path[<-] ($(foil1-1-2.south)-(0,1ex)$) edge[blue,bend left=-45] ($(foil1-1-8.south)-(0,1ex)$);
        \end{tikzpicture}
        \\[1ex] 
        \begin{tikzpicture}[>=stealth,baseline,anchor=base,inner sep=0pt]
            \matrix (foil2) [matrix of math nodes,nodes={minimum height=0.1em}] {
                & \ket{\downarrow}_{\mathcal{S}} & \otimes & {\color{red}\ket{\uparrow}_{\mathcal{E}}} &  &+ &  & \ket{\uparrow}_{\mathcal{S}} & \otimes & {\color{red}\ket{\downarrow}_{\mathcal{E}}} & \xrightarrow{~ {\color{red}U_{\mathcal{E}}}~} 
                \ket{\downarrow}_{\mathcal{S}}\otimes{\color{red}\ket{\downarrow}_{\mathcal{E}}}+\ket{\uparrow}_{\mathcal{S}}\otimes{\color{red}\ket{\uparrow}_{\mathcal{E}}}\,. \\
            };
            \path[->]  ($(foil2-1-4.north)+(0,1ex)$) edge[red,bend left=45] ($(foil2-1-10.north)+(0,1ex)$);
            \path[<-]  ($(foil2-1-4.south)-(0,1ex)$) edge[red,bend left=-45] ($(foil2-1-10.south)-(0,1ex)$);
        \end{tikzpicture}
    \end{array}
\end{equation}
A quantum state of a system $\mathcal{S}$ is envariant under a map $U_\mathcal{S}\otimes \id_\mathcal{E}$, if its action can be ``undone'' by a map on the ``environment'' $\mathcal{E}$, $\id_\mathcal{S}\otimes U_\mathcal{E}$. In this sense $\id_\mathcal{S}\otimes U_\mathcal{E}$ acts as the inverse of $U_\mathcal{S}\otimes\id_{\mathcal{E}}$.

Envariant states have played a central role in advancing our understanding of quantum foundations, including the Born's rule~\cite{Zurek2005PRA,schlosshauer2005on} and quantum-to-classical transitions~\cite{touil2024branching}. It has also been essential in building the foundations of statistical mechanics without having to rely on distinctly classical concepts, such as ensembles~\cite{deffner2016foundations,Zurek2018PR}. Moreover, simulations and experimental studies have deepened our understanding of its consequences~\cite{vermeyden2015experimental,harris2016quantum,deffner2017demonstration}.

Interestingly, it has also been shown that shortcuts to adiabaticity~\cite{STA2019RMP,Hatomura2024JPB}  can be understood from the perspective of entanglement~\cite{touil2021environment}. In Ref.~\cite{touil2021environment} some of us showed that excitations created locally in $\mathcal{S}$ can be ``undone'' by envariant maps acting only on $\mathcal{E}$. In other words, effectively adiabatic dynamics can be facilitated in open quantum systems by judiciously engineering the dynamics of the environment. This is of particular significance in the context of environment-assisted quantum control~\cite{touil2021environment,cooper2019environment,hansom2014environment,
xu2018environment,cimmarusti2015environment}, where an environment, or ancilla, is used as a resource to mediate operations on the target system. Usually, however, open system dynamics is not unitary. Thus, the natural question arises whether envariance is restricted to unitary operations, or whether similar symmetries exist in genuinely open systems.  If envariance can be maintained under general quantum maps, it would provide fundamental insight into system-environment correlations and offer a guiding principle for designing robust quantum control protocols for open systems.

However, to the very best of our knowledge, all previous analyses of envariance have been restricted to unitary operations. To further investigate this symmetry from the perspective of more general quantum operations (or simply ``operations''), we extend our framework beyond the unitary case to examine whether the envariance condition can be satisfied under more general circumstances. Specifically, we consider whether operations represented by generic, completely positive and trace-preserving (CPTP) maps~\cite{Nielsen}, $\Phi_{\mathcal{S}}$ and $\Phi_{\mathcal{E}}$, can achieve envariance.

Furthermore, from a fundamental point of view, our work offers deeper insight into the symmetry of entanglement underlying our quantum universe, with potential applications  including, e.g., quantum metrology~\cite{giovannetti2006quantum,lloyd2008enhanced,degen2016quantum}, quantum communication~\cite{siudzinska2020classical,lee2015classical,gyongyosi2018survey,hastings2009superadditivity, zhu2017superadditivity, zhuang2017additive}, quantum generative models~\cite{gao2022enhancing,sone2024quantum,huang2022quantum}, quantum error correction (QEC)~\cite{shor1995scheme, steane1996multiple, campbell2024series, bluvstein2024logical, google2023suppressing, fowler2012surface}, classical spacetime emergence via AdS/CFT correspondence~\cite{maldacena1999large,maldacena2003eternal,rangamani2017holographic,maldacena2013cool,van2010building,sahoo2020traversable,lantagne2020diagnosing,maldacena2018eternal,cottrell2019build,hirata2007ads,ryu2006holographic,ryu2006aspects,hawking2001desitter,bousso2024geometric,harlow2017ryu,
rocha2008evaporation,akers2020simple}, and many more.

This paper is organized as follows: In Sec.~\ref{sec:envariance}, we briefly review the concept of envariance and state the open problem. In Sec.~\ref{sec:preliminary_lemma}, we derive a supporting lemma stating that the local operations achieving envariance preserve the broader classes of correlations~\cite{hassan2013invariance}, such as quantum mutual information, classical correlations, and quantum discord~\cite{ollivier2001discord,henderson2001classical}. In Sec.~\ref{sec:main}, we first focus on bipartite systems and prove that to satisfy the envariance condition, the local operation must correspond to a Kraus operator-sum representation that achieves a decoherence-free subspace (DFS) ~\cite{lidar1998decoherence,lidar1999concatenating,kwiat2000experimental,bacon1999robustness,lidar2001decoherence,wang2013numerical,bacon2000universal} (See Theorem~\ref{theorem1}). Then, we show that  Theorem~\ref{theorem1} can be extended to multipartite cases (See Corollary~\ref{corollary}). In Sec.~\ref{sec:implications}, we then discuss the implications of our main theorems for environment-assisted shortcuts to adiabaticity and AdS/CFT. In Sec.~\ref{sec:example}, we provide an example of the main result, followed by the conclusion in Sec.~\ref{sec:conc}.

\section{Preliminaries -- envariance in closed and open dynamics}
\label{sec:envariance}

We start by establishing notions and notations. To this end, note that when $\mathcal{H}$ represents a Hilbert space, we use the superoperator $\Phi_{\mathcal{H}}(\cdot)$ to denote an operation acting on $\mathcal{H}$ and $\mathcal{I}_{\mathcal{H}}(\cdot)$ to represent the identity operation, respectively. Given a quantum state $\rho$, its von-Neumann entropy is defined as $S(\rho)\equiv-\Tr[\rho\ln\rho]$. Furthermore, we adopt natural units by setting the reduced Planck's constant $(\hbar)$ and Boltzmann's constant $(k_B)$ to unity, $\hbar = k_B = 1$.

\subsection{Review of envariance in unitary dynamics}
\label{sec:brief_review_envariance}
We provide a brief overview of the result by Paris in Ref.~\cite{paris2005unitary}. 
Consider a bipartite quantum system described by the tensor product of two Hilbert spaces, $\mathcal{S} \otimes \mathcal{E}$. Here, $\mathcal{S}$ and $\mathcal{E}$ represent the Hilbert spaces of the system of interest and that of its environment, respectively. Let the pure state $\ket{\psi} \in \mathcal{S} \otimes \mathcal{E}$ be expressed in the computational basis as
\begin{align}
\ket{\psi}=\sum_{i=1}^{\dim(\mathcal{S})}\sum_{j=1}^{\dim(\mathcal{E})}\psi_{ij}\ket{i}\otimes\ket{j}\,,
\label{eq:generic_pure}
\end{align}
where $\psi_{ij} \in \mathbb{C}$ denotes the $(i,j)$-th entry of a complex matrix $\Psi$ with dimensions $\dim(\mathcal{S}) \times \dim(\mathcal{E})$. Applying the singular value decomposition (SVD) to $\Psi$ yields $\Psi = \Omega_{\mathcal{S}}\, \Sigma\, \Omega_{\mathcal{E}}^{\top}$, where $\Sigma$ is a $\dim(\mathcal{S})\times\dim(\mathcal{E})$ rectangular diagonal matrix with $d$ strictly positive entries, given by $\Sigma_{ij} = \abs{c_i} \delta_{ij}$ for $1 \leq i \leq d$ and $\Sigma_{ij} = 0$ otherwise. These elements $\{\abs{c_k}\}_{k=1}^d$ represent the singular values of $\Psi$, lying within the interval $\abs{c_k} \in (0,1]$. The quantity $d = \rank(\Psi)$ will correspond to the Schmidt rank of $\ket{\psi}$ in Eq.~\eqref{eq:bipartite_entangle}. In addition, $\Omega_{\mathcal{S}}$ and $\Omega_{\mathcal{E}}$ are unitary matrices with dimensions $\dim(\mathcal{S}) \times \dim(\mathcal{S})$ and $\dim(\mathcal{E}) \times \dim(\mathcal{E})$, respectively.

The Schmidt decomposition of $\ket{\psi}$ takes the form
\begin{align}
    \ket{\psi} = \sum_{k=1}^{d} c_k \ket{s_k} \otimes \ket{e_k}\,.
    \label{eq:bipartite_entangle}
\end{align}
In this form, the complex numbers $c_k \in \mathbb{C}$ are known as the Schmidt coefficients, satisfying the normalization condition $\sum_{k=1}^{d} \abs{c_k}^2 = 1$. Here, $\{\ket{s_k}\}_{k=1}^{d}$ and $\{\ket{e_k}\}_{k=1}^{d}$ form orthonormal bases for the $d$-dimensional \textit{subspaces} $\mathcal{S}_d \subseteq \mathcal{S}$ and $\mathcal{E}_d \subseteq \mathcal{E}$, respectively. Also,  their orthogonal complements, $\overline{\mathcal{S}}_d \subseteq \mathcal{S}$ and $\overline{\mathcal{E}}_d \subseteq \mathcal{E}$, are spanned by $\{\ket{s_k}\}_{k=d+1}^{\dim(\mathcal{S})}$ and $\{\ket{e_k}\}_{k=d+1}^{\dim(\mathcal{E})}$, so that the full Hilbert spaces can be expressed as direct sums, i.e., $\mathcal{S} = \mathcal{S}_d \oplus \overline{\mathcal{S}}_d$ and $\mathcal{E} = \mathcal{E}_d \oplus \overline{\mathcal{E}}_d$. Note that the state $\ket{\psi}$ is a product state if and only if (abbreviated as ``iff'' from here on) $d=1$~\cite{Nielsen}.

The state $\ket{\psi}$ is called envariant under a local unitary $U_{\mathcal{S}}$ iff there exists a local unitary $U_{\mathcal{E}}$ which together with $U_{\mathcal{S}}$ leaves $\ket{\psi}$ invariant~\cite{Zurek2003PRL,Zurek2003RMP,Zurek2005PRA,Zurek2009NatPhys,Zurek2011PRL,zurek2022quantum}. Formally, the condition for the pure state $\ket{\psi}$ to be envariant under $U_{\mathcal{S}}$ is written as $\exists U_{\mathcal{E}}$ such that
\begin{align}
\ket{\psi}=\left(U_{\mathcal{S}}\otimes U_{\mathcal{E}}\right)\ket{\psi}\,.
\label{eq:envariance_unitary_relation}
\end{align}

In Ref.~\cite{paris2005unitary}, Paris rigorously analyzed the structure of the local unitaries on $\mathcal{S}$ and $\mathcal{E}$ that satisfy Eq.~\eqref{eq:envariance_unitary_relation}, and obtained the following characterization. From Eqs.~\eqref{eq:generic_pure} and \eqref{eq:envariance_unitary_relation}, the similarity transformations of $U_{\mathcal{S}}$ and $U_{\mathcal{E}}$ via $\Omega_{\mathcal{S}}$ and $\Omega_{\mathcal{E}}$, defined by 
\begin{align}
\begin{split}
R_{\mathcal{S}} &= \Omega_{\mathcal{S}}\ad U_{\mathcal{S}} \Omega_{\mathcal{S}}\\
R_{\mathcal{E}} &= \Omega_{\mathcal{E}}\ad U_{\mathcal{E}} \Omega_{\mathcal{E}}\,,
\end{split}
\end{align}
must satisfy the relation
\begin{align}
R_{\mathcal{S}}\Sigma = \Sigma R_{\mathcal{E}}^*\,.
\end{align}
where the symbol ``\,$*$\,'' denotes the complex conjugate. 
In particular, Paris derived the explicit structure of $R_{\mathcal{S}}$ and $R_{\mathcal{E}}$ as follows. Let $r_k$ denote the number of sets of $k$ identical Schmidt coefficients. For instance, $r_1$ is the number of distinct coefficients, $r_2$ counts the pairs, and so forth, satisfying $r_1 + 2r_2 + \cdots + n r_n = d$. Then, the unitary matrices decompose as
\begin{align}
    \begin{split}
        R_{\mathcal{S}} &= R_{\mathcal{S}_d} \oplus R_{\overline{\mathcal{S}}_d}\\
        R_{\mathcal{E}} &= R_{\mathcal{E}_d} \oplus R_{\overline{\mathcal{E}}_d}\,,
    \end{split}
\end{align} 
where $R_{\overline{\mathcal{S}}_d}$ and $R_{\overline{\mathcal{E}}_d}$ are arbitrary unitaries acting on the orthogonal complements of the support. The action within the support subspaces is given by
\begin{align}
\begin{split}
R_{\mathcal{S}_d} &= \left(\bigoplus_{k=1}^{r_1} e^{i\phi_k} \right) \oplus \left(\bigoplus_{k=1}^{r_2} V_k \right) \oplus \cdots \oplus \left(\bigoplus_{k=1}^{r_n} W_k \right) \\
R_{\mathcal{E}_d} &= \left(\bigoplus_{k=1}^{r_1} e^{-i\phi_k} \right) \oplus \left(\bigoplus_{k=1}^{r_2} V_k^{*} \right) \oplus \cdots \oplus \left(\bigoplus_{k=1}^{r_n} W_k^{*} \right)\,,
\end{split}
\label{eq:unitaryform}
\end{align}
where $\phi_1, \cdots, \phi_{r_1} \in \mathbb{R}$ are arbitrary phases, $V_1, \cdots, V_{r_2}$ are arbitrary $2 \times 2$ unitary matrices, and $W_1, \cdots, W_{r_n}$ are arbitrary $n \times n$ unitary matrices. These also ensure the uniqueness of $R_{\mathcal{E}_d}$ given a fixed $R_{\mathcal{S}_d}$.

\subsection{Envariance in non-unitary dynamics?}
\label{sec:open_problem}

However, the extension of the envariance condition to more general scenarios involving completely positive and trace-preserving (CPTP) maps beyond unitary operations has not been explored. A priori, it is unclear whether there exists a class of local CPTP maps,  $\Phi_{\mathcal{S}}$ and $\Phi_{\mathcal{E}}$,  that support an analogous envariance condition.

To address this question, we first need to formulate the envariance condition for a mixed state $\rho$. This, however, is rather straight forward if one recognizes that $(\Phi_{\mathcal{S}} \otimes \mathcal{I}_{\mathcal{E}})$ serves as the recovery map for $(\mathcal{I}_{\mathcal{S}} \otimes \Phi_{\mathcal{E}})$ and vice versa,  
\begin{align}
    \rho = \left((\Phi_{\mathcal{S}}\otimes\mathcal{I}_{\mathcal{E}})\circ(\mathcal{I}_{\mathcal{S}}\otimes\Phi_{\mathcal{E}})\right)(\rho) =(\Phi_{\mathcal{S}} \otimes \Phi_{\mathcal{E}})(\rho)\,.
\label{eq:envariance_condition_2body}
\end{align}
In this case, the reduced states,
\begin{align}
\begin{split}
\rho_{\mathcal{S}}&\equiv \Tr_{\mathcal{E}}\left[\rho\right]\\
\rho_{\mathcal{E}}&\equiv \Tr_{\mathcal{S}}\left[\rho\right]\,,
\end{split}
\label{eq:reduced}
\end{align}
are respectively the fixed points of $\Phi_{\mathcal{S}}$ and $\Phi_{\mathcal{E}}$, i.e. 
\begin{align}
\begin{split}
\Phi_{\mathcal{S}}(\rho_{\mathcal{S}})&=\rho_{\mathcal{S}}\\
\Phi_{\mathcal{E}}(\rho_{\mathcal{E}})&=\rho_{\mathcal{E}}\,.  
\end{split}
\label{eq:fixedpoint}
\end{align}
For simplicity, we refer to local operations $\Phi_{\mathcal{S}}$ and $\Phi_{\mathcal{E}}$ that satisfy the condition in Eq.~\eqref{eq:envariance_condition_2body} as \textit{envariance operations} for $\rho$~\footnote{The envariance operations considered here naturally include the identity operation, $\mathcal{I}_{\mathcal{S}} \otimes \mathcal{I}_{\mathcal{E}}$, as a trivial case. However, our primary focus is on the nontrivial cases. When referring to envariance operations in general, it implies that nontrivial cases are also included.}.

\subsection{Preliminary lemma}
\label{sec:preliminary_lemma}

In this section, we first show a useful preliminary lemma (Lemma~\ref{lemma1}).
Hassan and Joag~\cite{hassan2013invariance} proved that correlations in a bipartite quantum system remain invariant under local operations iff corresponding recovery operations exist for each subsystem. By applying their results, we immediately find that $(\Phi_{\mathcal{S}} \otimes \mathcal{I}_{\mathcal{E}})$ and $(\mathcal{I}_{\mathcal{S}} \otimes \Phi_{\mathcal{E}})$ preserve the quantum mutual information, classical correlations, and quantum discord (See Appendix~\ref{app:discord} for a brief review). Using the symbol $C(\rho)$ to collectively represent these correlations contained in $\rho$, we have 
\begin{align}
C(\rho)=C\left((\Phi_{\mathcal{S}}\otimes\mathcal{I}_{\mathcal{E}})(\rho)\right)=C\left((\mathcal{I}_{\mathcal{S}}\otimes\Phi_{\mathcal{E}})(\rho)\right)\,.
\end{align}

Particularly, when $\rho$ is a pure state $\rho=\dya{\psi}$, from Eq.~\eqref{eq:fixedpoint}, due to the invariance of the quantum mutual information of $\rho$ under $(\Phi_{\mathcal{S}}\otimes\mathcal{I}_{\mathcal{E}})$ or $(\mathcal{I}_{\mathcal{S}}\otimes\Phi_{\mathcal{E}})$, we find that the von-Neumann entropies of the intermediate states, 
\begin{align}
\begin{split}
\sigma &\equiv (\Phi_{\mathcal{S}} \otimes \mathcal{I}_{\mathcal{E}})(\rho)\\
\omega &\equiv (\mathcal{I}_{\mathcal{S}} \otimes \Phi_{\mathcal{E}})(\rho)\,,
\end{split}
\end{align}
become 
\begin{align}
S(\sigma)=S(\omega)=0\,.
\end{align}
We summarize this result in the following lemma (See Appendix~\ref{app:lemma} for the  proof):
\begin{lemma}
\label{lemma1}
\normalfont For a pure state $\rho=\dya{\psi}$, when $\Phi_{\mathcal{S}}$ and $\Phi_{\mathcal{E}}$ are envariance operations for $\rho$, the intermediate states,  $\sigma \equiv (\Phi_{\mathcal{S}} \otimes \mathcal{I}_{\mathcal{E}})(\rho)$ and $\omega \equiv (\mathcal{I}_{\mathcal{S}} \otimes \Phi_{\mathcal{E}})(\rho)$, are also pure.
\end{lemma}

\noindent Lemma~\ref{lemma1} implies that envariance, initially defined as the symmetry that preserves entanglement in pure states, corresponds to the preservation of correlations between quantum systems in the case of mixed states.

\section{No envariance for subspace non-unitary dynamics!}
\label{sec:main}

In this section, we present our main results, organized into a no-go theorem for bipartite systems, which also holds for the multipartite scenario.

\subsection{Bipartite systems}
\label{sec:bipartite}

Focusing on a bipartite system $\mathcal{S}\otimes\mathcal{E}$, in Theorem~\ref{theorem1}, we demonstrate that envariance operations correspond to those constructing the DFS. Then, Corollary~\ref{corollary} extends Theorem~\ref{theorem1} to the multipartite setting beyond the bipartite scenario.

Let us first consider a bipartite system with 
\begin{align}
d\leq\min\{\dim(\mathcal{S}),\dim(\mathcal{E})\}\,.
\end{align}
Suppose $\Phi_{\mathcal{S}}$ and $\Phi_{\mathcal{E}}$ are envariance operations for the pure state of  Eq.~\eqref{eq:bipartite_entangle},
\begin{align}
\rho=\dya{\psi}\,.
\end{align}
Let the Kraus operator-sum representation of $\Phi_{\mathcal{S}}$ be 
\begin{align}
\Phi_{\mathcal{S}}(\cdot) \equiv \sum_{\mu=1}^{M}\Gamma_{\mu}(\cdot)\Gamma_{\mu}\ad\,,
\end{align}
where $\{\Gamma_{\mu}\}_{\mu=1}^{M}$ are the Kraus operators acting on $\mathcal{S}$, which satisfy the completion relation 
\begin{align}
\sum_{\mu=1}^{M}\Gamma_{\mu}\ad \Gamma_{\mu}=\id_{\mathcal{S}}\,.
\end{align}
Here, $\id_{\mathcal{S}}$ is the identity operator acting on $\mathcal{S}$, and $M$ denotes the Kraus rank defined as the minimum number of Kraus operators needed to represent $\Phi_{\mathcal{S}}$~\cite{Nielsen}. From Lemma~\ref{lemma1},  $\sigma\equiv (\Phi_{\mathcal{S}}\otimes\mathcal{I}_{\mathcal{E}})(\rho)$
is a pure state; therefore, we have 
\begin{align}
    \sigma=\sigma^{2}\,.
\end{align} 
Then, we can prove that $\Phi_{\mathcal{S}}$ must act unitarily on the subspace $\mathcal{S}_d$, while potentially acting non-unitarily on their complementary subspaces $\overline{\mathcal{S}}_d$. Formally, for $\Phi_{\mathcal{S}}$, its Kraus operators $\Gamma_{\mu}$ must take the following direct-sum form:  
\begin{align}
\Gamma_{\mu} = (\sqrt{p_{\mu}}\,U_{\mathcal{S}_d}) \oplus \gamma_{\mu}\,,
\label{eq:DFS_S}
\end{align}
where the coefficients $p_{\mu}\in (0, 1]$ satisfy $\sum_{\mu=1}^{M} p_{\mu} = 1$, and $\{\gamma_{\mu}\}_{\mu=1}^{M}$ are Kraus operators acting on the complementary subspace $\overline{\mathcal{S}}_d$, which satisfies 
\begin{align}
\sum_{\mu=1}^{M}\gamma_{\mu}\ad\gamma_{\mu}=\id_{\overline{\mathcal{S}}_{d}}\,.
\end{align}
Taking into account that $\omega=(\mathcal{I}_{\mathcal{S}}\otimes\Phi_{\mathcal{E}})(\rho)$ is also a pure state, we prove that a similar structure holds for $\Phi_{\mathcal{E}}$, as well.

Note that $\Phi_{\mathcal{S}}$ and $\Phi_{\mathcal{E}}$ display the operator-sum representations necessary to build DFS~\cite{lidar1998decoherence,lidar1999concatenating,kwiat2000experimental,bacon1999robustness,lidar2001decoherence,wang2013numerical,bacon2000universal}, that is, a subspace of the Hilbert space evolves unitarily, even when the total Hilbert space undergoes noisy non-unitary processes. This condition provides a symmetry-protected method to cancel the effects of environmental noise. The DFS protocol is a passive method that protects quantum information by encoding it into subspaces that are immune to environmental noisy effects, in contrast to quantum error correction. We refer to the operations constructed by Kraus operators in the direct-sum form of Eq.~\eqref{eq:DFS_S} as \textit{DFS operations} and summarize the result in the following theorem (See Appendix~\ref{app:theorem1} for the  proof):
\begin{theorem}
\label{theorem1}
\normalfont When {$d\leq \min\{\dim(\mathcal{S}), \dim(\mathcal{E})\}$},  the envariance operations $\Phi_{\mathcal{S}}$ and $\Phi_{\mathcal{E}}$ for $\ket{\psi}$ must be DFS operations acting unitarily on the subspaces $\mathcal{S}_d$ and $\mathcal{E}_d$, respectively. 
\end{theorem}
As a main result, we have shown that the envariance condition can only be fulfilled by locally unitary maps. In other words, envariance is a symmetry unique to locally unitary dynamics~\footnote{{Extending our discussion into mixed states may lead to different result. As an illustrative example, let us consider a classically correlated state of the form $\rho=\sum_{k}\abs{c_k}^2\dya{s_k}\otimes\dya{e_k}$. In this case, the maps $\Phi_{\mathcal{S}}$ and $\Phi_{\mathcal{E}}$ are not required to act unitarily. Instead, they can be dephasing operations:  $\Phi_{\mathcal{S}}(\cdot)=\sum_{k}\dya{s_k}(\cdot)\dya{s_k}$ and  $\Phi_{\mathcal{E}}(\cdot)=\sum_{k}\dya{e_k}(\cdot)\dya{e_k}$, which results in $(\Phi_{\mathcal{S}}\otimes\Phi_{\mathcal{E}})(\rho)=\rho$.}}.

\subsection{Multipartite systems}
\label{sec:multipartite}

Next, let us consider a multipartite scenario where the system couples to multiple environments, 
\begin{align}  
\ket{\psi_n} = \sum_{k=1}^{d} c_k \ket{s_k} \otimes \ket{e_{k}^{(1)}, \cdots, e_{k}^{(n)}}\,,  
\label{eq:entangle_nbody} 
\end{align}  
where we defined 
$\ket{e_{k}^{(1)}, \cdots, e_{k}^{(n)}}\equiv\bigotimes_{j=1}^{n}\ket{e_{k}^{(j)}}$ and $n$ denotes the number of environments coupled to the system. Similar to the bipartite case, $\{\ket{e_{k}^{(j)}}\}_{k=1}^{d}$ form the complete orthonormal basis of the $d$--dimensional subspace $\mathcal{E}_d^{(j)}$ of $j$-th environment $\mathcal{E}^{(j)}$. Here, we can set
\begin{align}
{d\leq \min\left\{\dim(\mathcal{S}),\dim(\mathcal{E}^{(1)}),\cdots, \dim(\mathcal{E}^{(n)})\right\}\,.}
\label{eq:general_d}
\end{align}

Let us write
\begin{align}
\rho_n=\dya{\psi_n}\,.
\end{align}
When $\Phi_{\mathcal{S}},~\Phi_{\mathcal{E}^{(1)}},~\cdots,~\Phi_{\mathcal{E}^{(n)}}$ are the envariance operations for $\rho_n$, they satisfy 
\begin{align}
\left(\Phi_{\mathcal{S}}\otimes\Phi_{\mathcal{E}^{(1)}}\otimes\cdots\otimes \Phi_{\mathcal{E}^{(n)}}\right)(\rho_n)=\rho_n\,.
\end{align}
Then the reduced states of $\rho_n$,
\begin{align}
\begin{split}
&\rho_{\mathcal{S}}=\Tr_{\mathcal{E}^{(1)},\cdots,\mathcal{E}^{(n)}}[\rho_n]\\
&\rho_{\mathcal{E}^{(j)}}=\Tr_{\mathcal{S},\mathcal{E}^{(1)},\cdots,\mathcal{E}^{(j-1)},\mathcal{E}^{(j+1)},\cdots,\mathcal{E}^{(n)}}[\rho_n]
\end{split}
\end{align}
are again the fixed points of each operation
\begin{align}
\begin{split}
&\rho_{\mathcal{S}}=\Phi_{\mathcal{S}}(\rho_{\mathcal{S}})\\
&\rho_{\mathcal{E}^{(j)}}=\Phi_{\mathcal{E}^{(j)}}(\rho_{\mathcal{E}^{(j)}})\,. 
\end{split}
\end{align}
As we can see,  $(\mathcal{I}_{\mathcal{S}}\otimes\Phi_{\mathcal{E}^{(1)}}\otimes\cdots\otimes\Phi_{{\mathcal{E}}^{(n)}})$ is the recovery map for  $\Phi_{\mathcal{S}}\otimes\mathcal{I}_{\mathcal{E}^{(1)}}\otimes\cdots\otimes\mathcal{I}_{\mathcal{E}^{(n)}}$. Therefore, multipartite quantum mutual information defined as~\cite{Watanabe1960}
\begin{align}
I(\rho_n)\equiv S(\rho_{\mathcal{S}})+\sum_{j=1}^{n}S(\rho_{\mathcal{E}^{(j)}})-S(\rho_n)
\end{align}
is preserved under the operation $\Phi_{\mathcal{S}}\otimes\mathcal{I}_{\mathcal{E}^{(1)}}\otimes\cdots\otimes\mathcal{I}_{\mathcal{E}^{(n)}}$. Similar to the derivation of Lemma~\ref{lemma1}, this implies that the intermediate state 
\begin{align}
\sigma_n\equiv \left(\Phi_{\mathcal{S}}\otimes\mathcal{I}_{\mathcal{E}^{(1)}}\otimes\cdots\otimes\mathcal{I}_{\mathcal{E}^{(n)}}\right)(\rho_n)
\end{align}
is pure. This also holds for the operation $\Phi_{\mathcal{E}^{(j)}}$. Therefore, Theorem~\ref{theorem1} can be  extended to the multipartite case $(n\geq 3)$ as summarized in the following corollary:
\begin{corollary}
\label{corollary}
\normalfont When $d\leq\min\{\dim(\mathcal{S}),\dim(\mathcal{E}^{(1)}),\cdots,\dim(\mathcal{E}^{(n)})\}$, the envariance operations $\Phi_{\mathcal{S}}$ and $\Phi_{\mathcal{E}^{(j)}}~(j=1,2,\cdots,n)$ for $\ket{\psi_n}$ must be DFS operations acting unitarily on the subspaces $\mathcal{S}_d$ and $\mathcal{E}_{d}^{(j)}$, respectively. 
\end{corollary}
However, note that for the multipartite entanglement-assisted invariance, given a local unitary $U_{\mathcal{S}}$, the local unitary acting on the $j$-th environment $U_{\mathcal{E}^{(j)}}$ is \textit{not} unique as pointed out in Ref.~\cite{Zurek2005PRA}.

\section{Examples and no-go theorems}
\label{sec:implications}

We now continue by discussing immediate implications and consequences of our main results for shortcut to adiabaticity (STA) and AdS/CFT.

\subsection{Environment-assisted shortcuts to adiabaticity}
\label{sec:STA}

To determine the allowable operations for STA~\cite{campo2012assisted,campbell2015shortcut,demirplak2003adiabatic,demirplak2005assisted} to achieve the same dynamics described by the quantum adiabatic evolution~\cite{born1928beweis,kato1950adiabatic} from the perspective of envariance, we focus on the environment-assisted shortcuts to adiabaticity (EASTA), which was first introduced in Ref.~\cite{touil2021environment}. 

\begin{figure}[htp!]
\centering
\includegraphics[width=1\columnwidth]{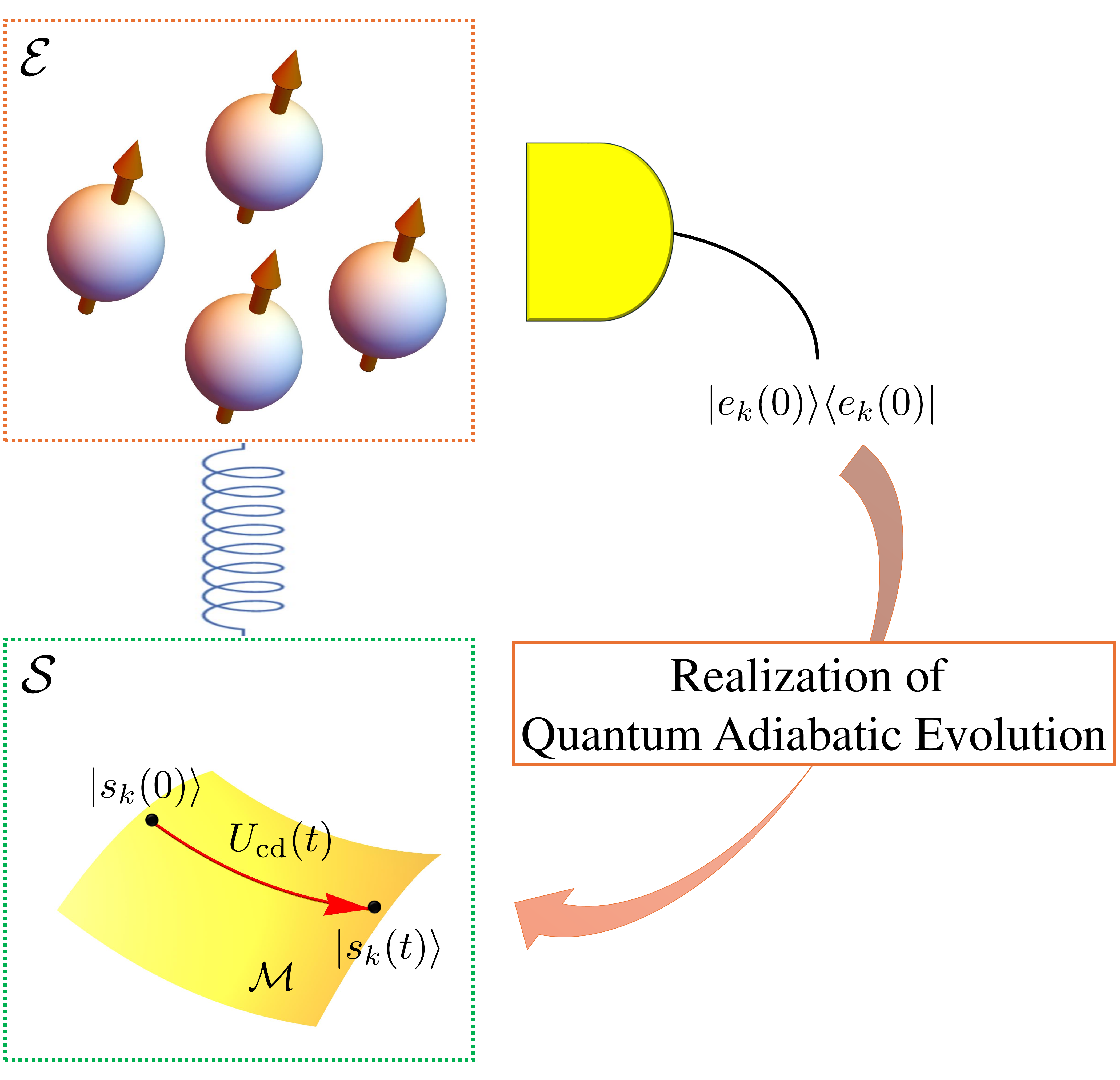}
\caption{Quantum adiabatic evolution of $\mathcal{S}$ through envariance. Assisted by the projection measurement $\ketbra{e_k(0)}{e_k(0)}$ on $\mathcal{E}$, the unitary $U_{\text{cd}}(t)$ generated by the counteradiabatic field enables the system to evolve through the adiabatic manifold $\mathcal{M}$ for all $t$.
}
\label{fig:adiabatic}
\end{figure}

By choosing $d = \dim(\mathcal{S}) = \dim(\mathcal{E})$, the initial state in EASTA is prepared as the maximally entangled state,  
\begin{align}
    \ket{\psi(0)}=\frac{1}{\sqrt{d}}\sum_{k=1}^{d}\ket{s_k(0)}\otimes\ket{e_k(0)}\,,
\label{eq:initial}
\end{align}
where $\{\ket{s_k(t)}\}_{k=1}^{d}$ are the eigenstates of the system Hamiltonian $H_{\mathcal{S}}(t)$, with instantaneous eigenvalues $\{E_k(t)\}_{k=1}^{d}$.  
Reference~\cite{touil2021environment} then demonstrated that for any given unitary $U_{\mathcal{S}}(t)$, there exists a unique unitary $U_{\mathcal{E}}(t)$ satisfying
\begin{align}
    (U_{\mathcal{S}}(t)\otimes U_{\mathcal{E}}(t))\ket{\psi(0)}
    =(U_{\rm{cd}}(t)\otimes\id_{\mathcal{E}})\ket{\psi(0)}\,.
\label{eq:EASTA}
\end{align}
Here, $U_{\rm{cd}}(t)$ generated by the counterdiabatic fields is defined as 
\begin{align}
U_{\rm{cd}}(t)\equiv\sum_{k=1}^{d}\ketbra{\varphi_k(t)}{s_k(0)}\,,
\end{align}
where we defined 
\begin{align}
\ket{\varphi_k(t)}=e^{-i\theta_k(t)}\ket{s_k(t)}~(\forall t)
\end{align}
with the time-dependent phase $\theta_k(t)$ given by 
\begin{align}
\theta_k(t)\equiv\int_{0}^{t}E_k(\tau)d\tau-i\int_{0}^{t}\bramatket{s_k(\tau)}{\partial_{\tau}}{s_{k}(\tau)}d\tau\,.
\end{align}
Therefore, $U_{\text{cd}}(t)$ enables $\mathcal{S}$ to evolve through the adiabatic manifold $\mathcal{M}$ for all $t$ (See Fig.~\ref{fig:adiabatic}),
and we can obtain the desired state $\ket{s_k(t)}$ by performing the projection measurement $\dya{e_k(0)}$ on $\mathcal{E}$. Then, the $(m,n)$th element of $U_{\mathcal{E}}(t)$ is given by ~\cite{touil2021environment}
\begin{align}
(U_{\mathcal{E}}(t))_{m,n} = \sum_{m,n}e^{-i\theta_{m}(t)}\bramatket{s_n(0)}{U_{\mathcal{S}}\ad (t)}{s_m(t)}\,,    
\end{align}
which can be derived from Eq.~\eqref{eq:unitaryform}. In EASTA, the main task is to construct $U_{\mathcal{E}}(t)$ for a given $U_{\mathcal{S}}(t)$.

The relation between the STA and EASTA can be interpreted as follows. If there exists a STA protocol, there always exists an EASTA protocol; therefore, the existence of STA protocol is a sufficient condition for the EASTA protocol. Then, its contraposition states if there is no EASTA protocol, then there is no corresponding STA protocol. This contraposition statement leads to the following no-go theorem for STA if we extend our scheme for EASTA to non-unitary operations. Our theorems state that only the locally unitary control can be used to achieve STA as 
\begin{align}
(U_{\rm{cd}}\ad(t)U_{\mathcal{S}}(t)\otimes U_{\mathcal{E}}(t))\ket{\psi(0)}=\ket{\psi(0)}\,.
\end{align}
Moreover, the unitary operator $ U_{\mathcal{E}}(t) $ is uniquely determined when $ U_{\rm{cd}}\ad(t) U_{\mathcal{S}}(t) $ is specified, which also confirms the result from Ref.~\cite{touil2021environment} regarding the uniqueness of $U_{\mathcal{E}}(t)$.

\subsection{AdS/CFT}
\label{sec:AdS/CFT}

The physics of cosmological black holes has found analogs in the properties of the charger carriers in strongly correlated quantum materials, such as graphene~\cite{
kandemir2017quasinormal, kandemir2020hairy, cvetivc2012graphene,chen2018quantumholography,franz2018mimicking}. By elucidating the implications of our main results within the AdS/CFT framework, we identify the requirements for environment-assisted quantum control to protect entanglement in two-dimensional quantum materials.

In the AdS/CFT correspondence, an eternal black hole in anti-de Sitter (AdS) spacetime is dual to an entangled state of two identical boundary conformal field theories (CFTs), expressed as
\begin{align}  
    \ket{\psi_{\beta}} \equiv \frac{1}{\sqrt{Z}} \sum_{k} e^{-\frac{\beta}{2}E_k} \ket{E_k}_L \otimes \ket{E_k}_R\,, 
\label{eq:TFD}
\end{align}  
known as the thermofield double (TFD) state of the composite system formed by the two CFTs~\cite{rangamani2017holographic,maldacena2013cool,maldacena2003eternal,van2010building,sahoo2020traversable,lantagne2020diagnosing,maldacena2018eternal,cottrell2019build,berenstein2019quenches, cai2020revisit}, conventionally referred to as ``right'' $(R)$ and ``left'' $(L)$ CFTs. {Here, we consider the black hole as capable of taking all possible energy eigenvalues, ensuring that $\ket{\psi_{\beta}}$ resides in the \textit{entire} bipartite Hilbert space.} This state is also the approximate ground state~\cite{sahoo2020traversable,lantagne2020diagnosing,maldacena2018eternal,cottrell2019build} of the Sachdev–Ye–Kitaev (SYK) model~\cite{sachdev1993gapless,kitaev2015syk1,kitaev2015syk2}.

The reduced states are given by the Gibbs state $\rho_L = \rho_R =  e^{-\beta H}/Z$, where $H \equiv \sum_{k} E_k \dya{E_k}$ is the Hamiltonian, with eigenvalues $E_k$ and corresponding eigenstates $\ket{E_k}$, and $Z \equiv \Tr[e^{-\beta H}]$ is the canonical partition function. Here, $\beta$ represents the inverse temperature of the black hole. In the context of AdS/CFT, $\ket{\psi_{\beta}}$ is the \textit{initial} state. Note that in this context, we focus on the asymptotically AdS regions, assuming that an observer in either region perceives the black hole spacetime, which is associated with the Gibbs state of a CFT~\cite{witten2001anti}. Additionally, the two CFTs are assumed to be \textit{non-interacting}, as the black hole horizons, which prevent communication between the two regions, naturally correspond to the absence of interactions between the CFTs~\cite{van2010building}. Therefore, only local operations on each CFT are considered natural in this context.

The black hole remains static under free evolution governed by the thermofield Hamiltonian $H_{\text{tot}}=H_L\otimes\id_{R}-\id_{L}\otimes H_R$, 
where $H_L=H_R=H$~\cite{maldacena2003eternal,maldacena2013cool,rangamani2017holographic}. By solving the Schr\"{o}dinger's equation, we immediately find that $\ket{\psi_{\beta}}$ remains invariant for all $t$, $\ket{\psi_{\beta}}=e^{-i H_{\text{tot}}t}\ket{\psi_{\beta}}=(e^{-i Ht}\otimes e^{+i H t})\ket{\psi_{\beta}}$, because the relative phases cancel each other.

\begin{figure}[htp!]
\centering
\includegraphics[width=.9\columnwidth]{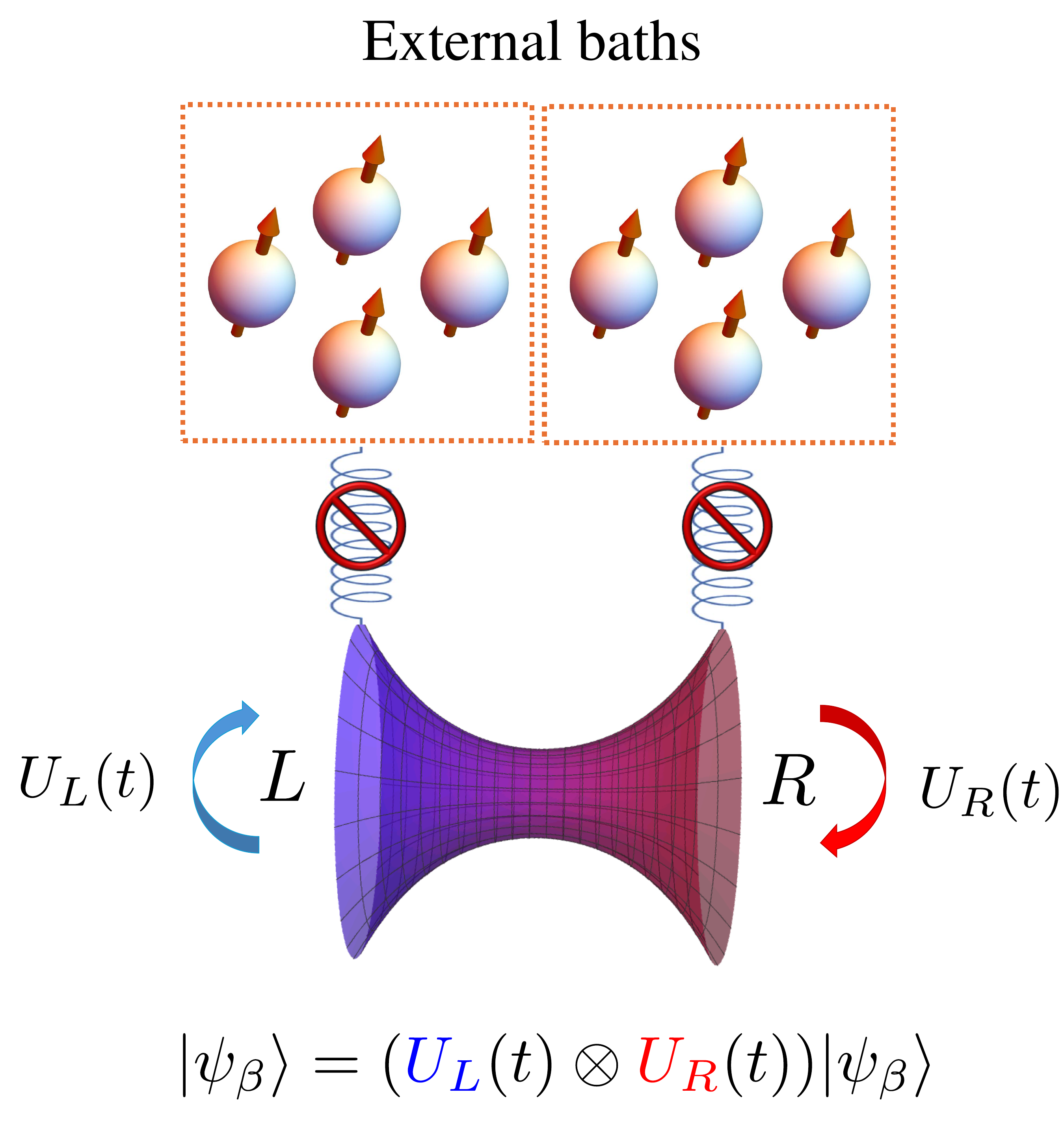}
\caption{Static condition of eternal black hole in AdS/CFT. The static condition {holds only for the closed CFTs,} realized {only} by local unitaries $U_L(t)$ and $U_R(t)$ with the forms of Eq.~\eqref{eq:unitaryform}. 
}
\label{fig:blackhole}
\end{figure}

Conversely, our theorems ensure that the black hole can be static only if the local operations applied to it are unitary. {Therefore, if we consider a scenario where the CFTs are coupled to the external baths (e.g. quantum many-body systems)~\cite{rocha2008evaporation,akers2020simple,tian2025dissipative, jana2020open, loganayagam2023holographic}, which causes the non-unitary evolution of the CFTs, the eternal black hole in AdS spacetime cannot be static due to the violation of the envariance symmetry. Therefore, the CFTs must be a closed system evolving under some local unitaries $U_L(t)$ and $U_R(t)$ satisfying Eq.~\eqref{eq:unitaryform} (See Fig.~\ref{fig:blackhole})}.
This also means that {for the closed CFTs} there could exist some other total Hamiltonians that leave the black hole system static beyond $H_{\text{tot}}$ even with time-dependence, {if the Hamiltonian generates $U_L(t)\otimes U_R(t)$.}
Furthermore, the generated $U_L(t)$ and $U_R(t)$ must satisfy the commutation relation
\begin{align}
\left[H,U_L(t)\right]=\left[H,U_R(t)\right]=0~(\forall t)\,,
\label{eq:staticcondition}
\end{align} 
given the fact that the Gibbs state must be the fixed point for the local unitaries.

\section{Example of DFS operations for envariance}
\label{sec:example}

Here, we provide an example of Theorem \ref{theorem1}. Consider the entangled state 
\begin{align}
    \ket{\psi} = \frac{1}{\sqrt{2}}\ket{0,0}_{\mathcal{S}} \otimes \ket{0,0}_{\mathcal{E}} + \frac{1}{\sqrt{2}}\ket{1,1}_{\mathcal{S}} \otimes \ket{1,1}_{\mathcal{E}}\,,
\end{align}
where we define $\ket{0} \equiv \begin{pmatrix} 1 & 0 \end{pmatrix}^T$ and $\ket{1} \equiv \begin{pmatrix} 0 & 1 \end{pmatrix}^T$. 
Here, $\mathcal{S}$ and $\mathcal{E}$ represent four-dimensional Hilbert spaces, and $\ket{\psi}$ resides in the subspace $\mathcal{S}_2 \otimes \mathcal{E}_2$, where $\mathcal{S}_2$ and $\mathcal{E}_2$ are spanned by the basis states $\ket{0,0}$ and $\ket{1,1}$.

An example of envariance operations for $\ket{\psi}$ is 
\begin{align}
    \Phi_{\mathcal{S}}(\cdot) = \Phi_{\mathcal{E}}(\cdot) = \Gamma_1 (\cdot) \Gamma_1\ad + \Gamma_2 (\cdot) \Gamma_2\ad\,,
\end{align}
where the Kraus operators are 
\begin{align}
\begin{split}
    \Gamma_1 =& \sqrt{p} \ketbra{0,0}{1,1} + \sqrt{p} \ketbra{1,1}{0,0}+ \sqrt{p} \ketbra{0,1}{1,0}\\
    =&
    \begin{pmatrix}
        0 & 0 & 0 & \sqrt{p} \\
        0 & 0 & \sqrt{p} & 0 \\
        0 & 0 & 0 & 0 \\
        \sqrt{p} & 0 & 0 & 0
    \end{pmatrix}
\end{split}
\end{align}
and
\begin{align}
\begin{split}
    \Gamma_2
    =& \sqrt{1-p} \ketbra{0,0}{1,1} + \sqrt{1-p} \ketbra{1,1}{0,0}\\
    &+ \dya{0,1} + \sqrt{1-p} \dya{1,0}\\
    =&\begin{pmatrix}
        0 & 0 & 0 & \sqrt{1-p} \\
        0 & 1 & 0 & 0 \\
        0 & 0 & \sqrt{1-p} & 0 \\
        \sqrt{1-p} & 0 & 0 & 0
    \end{pmatrix}
\end{split}
\end{align}
with $p \in (0,1)$. With this form, we find that 
\begin{align}
    (\Phi_{\mathcal{S}} \otimes \Phi_{\mathcal{E}})(\dya{\psi}) = \dya{\psi}\,.
\label{app:eq:recover}
\end{align}
We demonstrate that $\Phi_{\mathcal{S}}$ and $\Phi_{\mathcal{E}}$ are DFS operations for $\ket{\psi}$ as follows.

Define 
\begin{align}
    \ket{\vec{0}} \equiv \ket{0,0}\,,~
    \ket{\vec{1}} \equiv \ket{1,1}\,,~
    \ket{\vec{2}} \equiv \ket{0,1}\,,~
    \ket{\vec{3}} \equiv \ket{1,0}\,.
\end{align}
Then, we can write 
\begin{align}
    \ket{\psi} = \frac{1}{\sqrt{2}} \ket{\vec{0}}_{\mathcal{S}} \otimes \ket{\vec{0}}_{\mathcal{E}} + \frac{1}{\sqrt{2}} \ket{\vec{1}}_{\mathcal{S}} \otimes \ket{\vec{1}}_{\mathcal{E}}\,.
\end{align}
With this, we can also express 
\begin{align}
\begin{split}
    \Gamma_1 &= \sqrt{p} (\ketbra{\vec{0}}{\vec{1}} + \ketbra{\vec{1}}{\vec{0}}) + \sqrt{p} \ketbra{\vec{2}}{\vec{3}}\\ 
    \Gamma_2 &= \sqrt{1-p} (\ketbra{\vec{0}}{\vec{1}} + \ketbra{\vec{1}}{\vec{0}}) + \dya{\vec{2}} + \sqrt{1-p} \dya{\vec{3}}\,.
\end{split}
\end{align}
In the subspace spanned by $\ket{\vec{0}}$ and $\ket{\vec{1}}$, $(\ketbra{\vec{0}}{\vec{1}} + \ketbra{\vec{1}}{\vec{0}})$ corresponds to the Pauli-$X$ gate, the quantum equivalent of a NOT gate. Let us write 
\begin{align}
    X \equiv \ketbra{\vec{0}}{\vec{1}} + \ketbra{\vec{1}}{\vec{0}}\,.
\end{align}
In the complementary subspace spanned by $\ket{\vec{2}}$ and $\ket{\vec{3}}$, defining 
\begin{align}
\begin{split}
    &\gamma_1 \equiv \begin{pmatrix} 0 & \sqrt{p} \\ 0 & 0 \end{pmatrix}\\
    &\gamma_2 \equiv \begin{pmatrix} 1 & 0 \\ 0 & \sqrt{1-p} \end{pmatrix}\,,
\end{split}
\end{align}
we can write 
\begin{align}
\begin{split}
    \Gamma_1 &= (\sqrt{p}\,X) \oplus \gamma_1 = \begin{pmatrix} \sqrt{p}\,X &  \\  & \gamma_1 \end{pmatrix}\\
    \Gamma_2 &= (\sqrt{1-p}\,X) \oplus \gamma_2 = \begin{pmatrix} \sqrt{1-p}\,X &  \\  & \gamma_2 \end{pmatrix}\,.
\end{split}
\end{align}
Here, $\gamma_1$ and $\gamma_2$ are Kraus operators describing amplitude damping~\cite{Nielsen} in the subspace spanned by $\ket{\vec{2}}$ and $\ket{\vec{3}}$.
Thus, $\Phi_{\mathcal{S}}$ and $\Phi_{\mathcal{E}}$ act on the subspace $\mathcal{S}_2$ and $\mathcal{E}_2$ with $X$ while causing dissipation in their complementary subspaces.

Since $\ket{\psi}$ resides in the subspace spanned by $\ket{\vec{0},\vec{0}}$, $\ket{\vec{0},\vec{1}}$, $\ket{\vec{1},\vec{0}}$, and $\ket{\vec{1},\vec{1}}$, the operators $\gamma_1$ and $\gamma_2$ do \textit{not} affect this subspace. Consequently, $\ket{\psi}$ does not undergo the dissipative process. At the same time, $\ket{\psi}$ remains envariant under the $X$ operation as
\begin{equation}
    \begin{array}{c}
         \begin{tikzpicture}[>=stealth,baseline,anchor=base,inner sep=0pt]
            \matrix (foil1) [matrix of math nodes,nodes={minimum height=0.1em}] {
                & {\color{blue}\ket{\vec{0}}_{\mathcal{S}}} & \otimes & \ket{\vec{0}}_{\mathcal{E}} &  & + &  & {\color{blue}\ket{\vec{1}}_{\mathcal{S}}} & \otimes & \ket{\vec{1}}_{\mathcal{E}} &  \xrightarrow{~ {\color{blue}X}~} 
                {\color{blue}\ket{\vec{1}}_{\mathcal{S}}}\otimes\ket{\vec{0}}_{\mathcal{E}}+{\color{blue}\ket{\vec{0}}_{\mathcal{S}}}\otimes\ket{\vec{1}}_{\mathcal{E}}\\
            };
            \path[->] ($(foil1-1-2.north)+(0,1ex)$) edge[blue,bend left=45] ($(foil1-1-8.north)+(0,1ex)$);
            \path[<-] ($(foil1-1-2.south)-(0,1ex)$) edge[blue,bend left=-45] ($(foil1-1-8.south)-(0,1ex)$);
        \end{tikzpicture}
        \\[1ex] 
        \begin{tikzpicture}[>=stealth,baseline,anchor=base,inner sep=0pt]
            \matrix (foil2) [matrix of math nodes,nodes={minimum height=0.1em}] {
                & \ket{\vec{1}}_{\mathcal{S}} & \otimes & {\color{red}\ket{\vec{0}}_{\mathcal{E}}} &  &+ &  & \ket{\vec{0}}_{\mathcal{S}} & \otimes & {\color{red}\ket{\vec{1}}_{\mathcal{E}}} & \xrightarrow{~ {\color{red}X}~} 
                \ket{\vec{1}}_{\mathcal{S}}\otimes{\color{red}\ket{\vec{1}}_{\mathcal{E}}}+\ket{\vec{0}}_{\mathcal{S}}\otimes{\color{red}\ket{\vec{0}}_{\mathcal{E}}}\,. \\
            };
            \path[->]  ($(foil2-1-4.north)+(0,1ex)$) edge[red,bend left=45] ($(foil2-1-10.north)+(0,1ex)$);
            \path[<-]  ($(foil2-1-4.south)-(0,1ex)$) edge[red,bend left=-45] ($(foil2-1-10.south)-(0,1ex)$);
        \end{tikzpicture}
    \end{array}
\end{equation}
From this, $\Phi_{\mathcal{S}}$ and $\Phi_{\mathcal{E}}$ are the DFS operations and  Eq.~\eqref{app:eq:recover} holds.

\section{Concluding remarks}
\label{sec:conc}

In this analysis, we have proven that envariance requires a Kraus operator sum representation encompassing a decoherence-free subspace (Theorem~\ref{theorem1}). Notably, this can be generalized to multipartite systems beyond the bipartite case (Corollary~\ref{corollary}). As a main result, we have thus shown that envariance is a symmetry unique to locally unitary operations.

Our findings have several immediate and important implications. We have demonstrated that extending counterdiabatic driving  to non-unitary dynamics is not feasible. Moreover, in the AdS/CFT context, we have shown that an eternal black hole in AdS spacetime cannot remain static with if the CFTs are coupled to the external baths. This also provides insights into protecting entanglement in strongly correlated quantum materials, where black hole physics naturally emerges.

Our results clarify the operational requirements for preserving envariance and provide insight into the structural properties of entanglement in quantum systems. By refining the set of permissible operations, our work provides a robust framework for understanding the role of envariance and operations in preserving entanglement across various systems.

\section*{Acknowledgement}
We would like to offer our gratitude to Wojciech H. Zurek and Bin Yan for helpful feedback on the main results of the paper. A.S. and P.C. acknowledge U.S. NSF under Grant No. OSI-2328774. A.S. also acknowledges PHY-2425180 and Cooperative Agreement PHY-2019786. A.T. is supported by the U.S DOE under the LDRD program at Los Alamos. K.M. is supported by the Goldwater scholarship and CSM Undergraduate Research Fellowship at UMass Boston.  S.D. acknowledges support from the John Templeton Foundation under Grant No. 62422.

\appendix

\section{Review of correlations in quantum systems}
\label{app:discord}
For a given mixed state $\rho$ in the composite system $\mathcal{S} \otimes \mathcal{E}$, the reduced states are defined as 
\begin{align}
\begin{split}
\rho_{\mathcal{S}}&\equiv \Tr_{\mathcal{E}}[\rho]\\
\rho_{\mathcal{E}}&\equiv \Tr_{\mathcal{S}}[\rho]\,.
\end{split}
\label{app:eq:reduced}
\end{align}
Let $\{A_a\}_{a=1}^{N_{\mathcal{S}}}$ be the set of positive operator-valued measurements (POVMs) performed on the subsystem $\mathcal{S}$, which satisfies the following completion relation 
\begin{align}
\sum_{a=1}^{N_{\mathcal{S}}} A_a\ad A_a=\id_{\mathcal{S}}\,.
\end{align}
The quantum mutual information~\cite{Nielsen} contained in $\rho$ is given by
\begin{align}
I(\rho)\equiv S(\rho_{\mathcal{S}}) + S(\rho_{\mathcal{E}}) - S(\rho)\,.
\end{align}

The quantum discord induced by the complete set of POVMs $\{A_a\}_{a=1}^{N_{\mathcal{S}}}$ is defined as~\cite{ollivier2001discord,henderson2001classical}
\begin{align}
D_{\mathcal{S} \to \mathcal{E}}(\rho) \equiv I(\rho)-J_{\mathcal{S}\to \mathcal{E}}(\rho)\,,
\end{align} 
where 
\begin{align}
J_{\mathcal{S} \to \mathcal{E}}(\rho)\equiv S(\rho_{\mathcal{E}})-\min_{\{A_a\}}\sum_{a=1}^{N_{\mathcal{S}}}p_a S(\rho_{\mathcal{E}|A_a})
\end{align}
denotes the classical correlations as a result of the POVMs. Here, $p_a$ is the probability defined as 
\begin{align}
p_a\equiv \Tr\left[(A_a\ad A_a\otimes\id_{\mathcal{E}})\rho\right] = \Tr\left[A_a\ad A_a\rho_{\mathcal{S}}\right]\,,
\end{align}
and $\rho_{\mathcal{E}|A_a}$ denotes the post-measurement state defined as  
\begin{align}
\rho_{\mathcal{E}|A_a}\equiv \frac{(A_a\otimes\id_{\mathcal{E}})\,\rho\,(A_a\ad\otimes\id_{\mathcal{E}})}{p_a}\,.    
\end{align}

Similarly, we can also construct the classical correlations $J_{\mathcal{E}\to \mathcal{S}}(\rho)$ and quantum discord $D_{\mathcal{E}\to \mathcal{S}}(\rho)\equiv I(\rho)-J_{\mathcal{E}\to \mathcal{S}}(\rho)$ by considering the complete set of POVMs $\{B_{b}\}_{b=1}^{N_{\mathcal{E}}}$ on the subsystem $\mathcal{E}$. Note that in general, $D_{\mathcal{S}\to \mathcal{E}}(\rho)$ is not necessarily equivalent to $D_{\mathcal{E}\to \mathcal{S}}(\rho)$.

\section{Proof of Lemma \ref{lemma1}}
\label{app:lemma}
When $\Phi_{\mathcal{S}}$ is an envariance operation for a pure state $\rho = \dya{\psi}$, the operation $(\Phi_{\mathcal{S}} \otimes \mathcal{I}_{\mathcal{E}})$ and $(\mathcal{I}_{\mathcal{S}}\otimes\Phi_{\mathcal{E}})$ preserve the quantum mutual information,  
\begin{align}  
\begin{split}
S(\rho_{\mathcal{S}}) &+ S(\rho_{\mathcal{E}}) - S(\rho)\\
&= S(\Phi_{\mathcal{S}}(\rho_{\mathcal{S}})) + S(\rho_{\mathcal{E}}) - S\left((\Phi_{\mathcal{S}} \otimes \mathcal{I}_{\mathcal{E}})(\rho)\right)\\
&= S(\rho_{\mathcal{S}}) + S(\Phi_{\mathcal{E}}(\rho_{\mathcal{E}})) - S\left((\mathcal{I}_{\mathcal{S}} \otimes \Phi_{\mathcal{E}})(\rho)\right)\,.
\end{split}
\end{align}
From
\begin{align}
\begin{split}
 \rho_{\mathcal{S}}&=\Phi_{\mathcal{S}}(\rho_{\mathcal{S}})\\
 \rho_{\mathcal{E}}&=\Phi_{\mathcal{E}}(\rho_{\mathcal{E}})
\end{split}
\label{app:eq:fixed}
\end{align}
and 
\begin{align}
S(\rho) = S(\dya{\psi}) = 0\,,
\end{align}
we obtain 
\begin{align}
S\left((\Phi_{\mathcal{S}} \otimes \mathcal{I}_{\mathcal{E}})(\rho)\right) =
S\left((\mathcal{I}_{\mathcal{S}} \otimes \Phi_{\mathcal{E}})(\rho)\right) = 0\,.
\end{align}
Therefore, the intermediate states,   $\sigma \equiv (\Phi_{\mathcal{S}} \otimes \mathcal{I}_{\mathcal{E}})(\rho)$ and $\omega\equiv (\mathcal{I}_{\mathcal{S}} \otimes \Phi_{\mathcal{E}})(\rho)$, are also pure states.

\section{Proof of Theorem \ref{theorem1}}
\label{app:theorem1}

Let us consider a pure entangled state $\rho=\dya{\psi}$ with
\begin{align}
    \ket{\psi}=\sum_{k=1}^{d}c_k\ket{s_k}\otimes\ket{e_k}\,.
\label{app:eq:entangled}
\end{align}
In this case, the reduced states are 
\begin{align}
\begin{split}
\rho_{\mathcal{S}}&\equiv\Tr_{\mathcal{E}}[\rho]=\sum_{k=1}^{d}\abs{c_k}^2\dya{s_k}\\
\rho_{\mathcal{E}}&\equiv\Tr_{\mathcal{S}}[\rho]=\sum_{k=1}^{d}\abs{c_k}^2\dya{e_k}\,.
\end{split}
\label{app:eq:reduced}
\end{align}
Let $\Phi_{\mathcal{S}}$ and $\Phi_{\mathcal{E}}$ be envariance operations for $\ket{\psi}$.  Since $(\Phi_{\mathcal{S}}\otimes\mathcal{I}_{\mathcal{E}})(\rho)$ is a pure state, defining 
\begin{align}
\begin{split}
    \sigma &\equiv (\Phi_{\mathcal{S}}\otimes\mathcal{I}_{\mathcal{E}})(\rho)\\
    &=\sum_{k,\ell}c_kc_{\ell}^{*}\Phi_{\mathcal{S}}(\ketbra{s_k}{s_{\ell}})\otimes \ketbra{e_k}{e_{\ell}}\,,
\end{split}
\label{app:eq:rho_S_2body}
\end{align}
we must have 
\begin{align}
    \sigma = \sigma^{2}\,.
\label{app:eq:purity2body}
\end{align}
From Eq.~\eqref{app:eq:rho_S_2body}, we can obtain 
\begin{widetext}
\begin{align}
\sigma^{2}=\sum_{k,\ell} c_k c_{\ell}^{*} \left(\sum_{j=1}^{d}\abs{c_{j}}^2 \Phi_{\mathcal{S}}(\ketbra{s_k}{s_j})\Phi_{\mathcal{S}}(\ketbra{s_j}{s_{\ell}})\right)\otimes\ketbra{e_k}{e_{\ell}}\,.
\label{app:eq:sigma^2}
\end{align}
\end{widetext}
\noindent Because $\{\ket{e}_k\}_{k=1}^{d}$ form the complete orthonormal basis of  {the subspace $\mathcal{E}_d$}, from Eqs.~\eqref{app:eq:rho_S_2body}, \eqref{app:eq:purity2body} and \eqref{app:eq:sigma^2}, we must have 
\begin{align}
    \Phi_{\mathcal{S}}(\ketbra{s_k}{s_{\ell}}) = \sum_{j=1}^{d}\abs{c_j}^2\Phi_{\mathcal{S}}(\ketbra{s_k}{s_j})\Phi_{\mathcal{S}}(\ketbra{s_j}{s_{\ell}})\,.
\label{app:eq:element_2body}
\end{align}

Now, let us use the Kraus operator-sum representation for $\Phi_{\mathcal{S}}$ by writing
\begin{align}
    \Phi_{\mathcal{S}}(\cdot)\equiv \sum_{\mu=1}^{M}\Gamma_{\mu}(\cdot)\Gamma_{\mu}\ad\,,
\label{app:eq:KrausOperatorSumSd}
\end{align}
where $\{\Gamma_{\mu}\}_{\mu=1}^{M}$ denotes the complete set of Kraus operators acting on $\mathcal{S}$ and $M$ is the Kraus rank, which is defined as the minimum number of Kraus operators required to represent the corresponding operation $\Phi_{\mathcal{S}}$~\cite{Nielsen}. Because $\Phi_{\mathcal{S}}$ is an operation, the Kraus operators must satisfy
\begin{align}
    \sum_{\mu=1}^{M}\Gamma_{\mu}\ad \Gamma_{\mu}=\id_{\mathcal{S}}\,.
\label{app:eq:closure}
\end{align}
Then, the both sides of Eq.~\eqref{app:eq:element_2body} can be explicitly written as 
\begin{align}
    \Phi_{\mathcal{S}}(\ketbra{s_k}{s_{\ell}}) = \sum_{\mu=1}^{M}\Gamma_{\mu}\ketbra{s_k}{s_{\ell}}\Gamma_{\mu}\ad
\label{app:eq:Kraus_1_2body}
\end{align}
and
\begin{align}
    \begin{split}
        \sum_{j=1}^{d}& \abs{c_j}^2\Phi_{\mathcal{S}}(\ketbra{s_k}{s_j})\Phi_{\mathcal{S}}(\ketbra{s_j}{s_{\ell}})\\
        &=\sum_{\mu,\nu}\left(\sum_{j=1}^{d}\abs{c_j}^2\bramatket{s_j}{\Gamma_{\mu}\ad \Gamma_{\nu}}{s_j}\right)\Gamma_{\mu}\ketbra{s_k}{s_{\ell}}\Gamma_{\nu}\ad\\
        &=\sum_{\mu,\nu}\Tr\left[\rho_{\mathcal{S}}\Gamma_{\mu}\ad \Gamma_{\nu}\right]\Gamma_{\mu}\ketbra{s_k}{s_{\ell}}\Gamma_{\nu}\ad\,.
    \end{split}
\label{app:eq:Kraus_2_2body}
\end{align}
Since Eq.~\eqref{app:eq:element_2body} has to hold for all $\ket{s_k}$, from Eqs.~\eqref{app:eq:Kraus_1_2body} and \eqref{app:eq:Kraus_2_2body}, we need 
\begin{align}
    \Tr\left[\rho_{\mathcal{S}}\Gamma_{\mu}\ad \Gamma_{\nu}\right]=\delta_{\mu\nu}\equiv\begin{cases}
        1~~(\mu=\nu)\\
        0~~(\mu\neq\nu)
    \end{cases}\,,
\label{app:eq:KroneckerDelta1}
\end{align}
which leads to
\begin{align}
\Tr\left[\rho_{\mathcal{S}}\Gamma_{\mu}\ad \Gamma_{\mu}\right]=1\,.
\label{app:eq:trace_unitary}
\end{align}
Also, from Eq.~\eqref{app:eq:closure}, we have 
\begin{align}
    \sum_{\mu=1}^{M}\Tr\left[\rho_{\mathcal{S}}\Gamma_{\mu}\ad \Gamma_{\mu}\right]=1\,.
\label{app:eq:trace_kraus_2body}
\end{align}
Therefore, from Eqs.~\eqref{app:eq:trace_unitary} and \eqref{app:eq:trace_kraus_2body}, the Kraus rank must be 
\begin{align}
    M = 1\,.
\label{app:eq:M=1}
\end{align}
This means that the minimum number of Kraus operators needed to represent the quantum operation acting on the subspace $\mathcal{S}_{d}=\text{Span}\left(\{\ket{s_j}\}_{j=1}^{d}\right)$, is 1. 
This implies that $\Phi_{\mathcal{S}}$ must act unitarily on the subspace $\mathcal{S}$, while it may act non-unitarily on $\overline{\mathcal{S}}_d$. This is precisely the property of the quantum process that establishes a decoherence-free subspace (DFS)~\cite{lidar1998decoherence,lidar1999concatenating,kwiat2000experimental,bacon1999robustness,lidar2001decoherence,wang2013numerical,bacon2000universal}. Formally, each Kraus operator $\Gamma_{\mu}$ must have the following direct-sum form: 
\begin{align}
    \Gamma_{\mu} = (\sqrt{p_{\mu}}\,U_{\mathcal{S}_d}) \oplus \gamma_{\mu} = 
    \begin{pmatrix}
        \sqrt{p_{\mu}}\,U_{\mathcal{S}_d}&\\
        &\gamma_{\mu}
    \end{pmatrix},
\end{align}
where $p_{\mu}\in(0, 1]~(\forall \mu)$ satisfies $\sum_{\mu=1}^{M}p_{\mu} = 1$ and $\{\gamma_{\mu}\}_{\mu=1}^{M}$ is the set of Kraus operators acting on $\overline{\mathcal{S}}_d$. Therefore, $\Phi_{\mathcal{S}}$ is a DFS operation. In a same approach, we can also prove that $\Phi_{\mathcal{E}}$ must be a DFS operation as well.

\bibliography{ref.bib}

\end{document}